# Machine Learning-Powered Data Cleaning for LEGEND: A Semi-Supervised Approach Using Affinity Propagation and Support Vector Machines

E. León[1,2,3], A. Li[3,4], M.A. Bahena Schott[1], B. Bos[1,2],
M. Busch[2,5], J.R. Chapman[1,2], G.L. Duran[1,2], J. Gruszko[1,2],
R. Henning[1,2], E.L. Martin[2,5], J.F. Wilkerson[1,2]

[1] Department of Physics and Astronomy, University of North Carolina, Chapel Hill, NC 27599, USA
[2] Triangle Universities Nuclear Laboratory, Durham, NC 27708, USA
[2] Data Science & Statistical Modeling, Analysis Group, Boston, MA 02119, United States of America
[3] Department of Physics, UC San Diego, La Jolla, CA, 92093, USA
[4] Halıcıoğlu Data Science Institute, UC San Diego, La Jolla, CA 92093, USA
[5] Department of Physics, Duke University, Durham, NC 27708, USA

E-mail: esleon97@gmail.com

6 April 2025

**Abstract.** Neutrinoless double-beta decay ($0\nu\beta\beta$) is a rare nuclear process that, if observed, will provide insight into the nature of neutrinos and help explain the matter-antimatter asymmetry in the universe. The Large Enriched Germanium Experiment for Neutrinoless Double-Beta Decay (LEGEND) will operate in two phases to search for $0\nu\beta\beta$. The first (second) stage will employ 200 (1000) kg of High-Purity Germanium (HPGe) enriched in $^{76}$Ge to achieve a half-life sensitivity of $10^{27}$ ($10^{28}$) years. In this study, we present a semi-supervised data-driven approach to remove non-physical events captured by HPGe detectors powered by a novel artificial intelligence model. We utilize Affinity Propagation to cluster waveform signals based on their shape and a Support Vector Machine to classify them into different categories. We train, optimize, and test our model on data taken from a natural abundance HPGe detector installed in the Full Chain Test experimental stand at the University of North Carolina at Chapel Hill. We demonstrate that our model yields a maximum sacrifice of physics events of $0.024^{+0.004}_{-0.003}\%$ after data cleaning. Our model is being used to accelerate data cleaning development for LEGEND-200 and will serve to improve data cleaning procedures for LEGEND-1000.

## 1. Introduction

The Large Enriched Germanium Experiment for Neutrinoless Double-Beta Decay (LEGEND) [1] is a large-scale international experiment that uses a phased approach to discover neutrinoless double-beta decay ($0\nu\beta\beta$) [2] using high purity Germanium detectors (HPGe). LEGEND combines the best technologies from the previous germanium-based experiments, namely, the GErmanium Detector Array (GERDA) [3] and the MAJORANA DEMONSTRATOR (MJD) [4].

Signals captured by HPGe detectors pass through an amplifying electronics chain before being digitized and saved to memory. The digitized signals are also referred to as waveforms. Since LEGEND operates in a low-background environment, a considerable fraction of the recorded data corresponds to non-physical waveforms caused by electronic noise and transient anomalies in the data acquisition (DAQ) system. In order to analyze the data, these anomalous events must be tagged during digital signal processing. This process is referred to as data cleaning.

Traditional data cleaning methods rely on procedures in which the scientist must browse through a comprehensive sample of the data to find all the existing types of anomalous events. The scientist must then develop parameters that can discriminate anomalous events, and perform cuts based on these parameters to tag said anomalous events. These parameters can vary over time and by detector. LEGEND-200 will run for five years and with four detector types: p-type point-contact detectors made by ORTEC® (PPC) [5], Broad Energy Germanium (BEGe) made by Mirion®[6], Inverted Coaxial Point-Contact (ICPC) detectors made by ORTEC ® and Mirion® [7, 8], and semi-coaxial (COAX) detectors refurbished from previous experiments. In some cases, the different detector geometries require dedicated traditional data cleaning cuts. Different hardware configurations and run conditions, such as detector characterization systems [9], test stands [10], or commissioning runs, affect the performance and stability of these parameters. For example, in traditional data cleaning for LEGEND-200, 12 individual digital signal processing parameters, each with detector-specific acceptance ranges, are used to construct four event categories. Different run periods of the experiment require different parameter acceptance ranges, and in some cases, newly-developed parameters. The traditional data cleaning methods used in previous-generation HPGe-based experiments are detailed in [11, 12]. Overall, data cleaning with traditional procedures requires a significant amount of time and human effort.

Consequently, we propose a data cleaning mechanism based on machine learning (ML). In the experimental searches for $0\nu\beta\beta$-decay, ML has proven useful to distinguish signal from background events [13, 14, 15], classify events based on spatiotemporal information [16], study inactive regions of detectors [17], reject pileup events [18], and match simulations to data [19]. Our model is based on two ML algorithms: Affinity Propagation (AP) [20] and Support Vector Machine (SVM) [21]. AP is an unsupervised learning clustering algorithm that groups signals in our datasets based on their pulse shape and assigns them to a cluster with a corresponding label. AP also provides the ability to automatically identify new event clusters directly from the data as run conditions change over time, serving as a form of anomaly detection. We re-group the cluster labels assigned by AP in terms of data cleaning categories. SVM is a supervised learning classifier which takes in signals and separates them based on the labels provided by the user.

By training our model on a comprehensive subset of the data, we can expand its predictive power to classify events in larger datasets in a data-driven and autonomous manner. Our model has been applied to a variety of different run configurations and setups containing HPGe detectors, proving its flexibility. For this study, we evaluate the performance of our model on data taken from the Full Chain Test (FCT) experimental stand at the University of North Carolina at Chapel Hill. The FCT, which replicates the full LEGEND-200 HPGe electronics chain in a liquid Argon test stand, was assembled to study the performance of the detectors, electronics, and digitizers prior to initial detector deployment and provide rapid feedback during operations.

In Section 2 we describe the experimental setup of the FCT. Section 3 provides details on our ML data cleaning model. Section 4 summarizes the training and optimization process of the model. In section 5 we test the effectiveness of our technique via sacrifice and leakage studies. Section 6 provides concluding remarks. The code and sample data sets needed to reproduce this work are publicly available at [22] and [23] respectively.

## 2. Experimental setup: Full Chain Test

The Full Chain Test (FCT) was built at the University of North Carolina at Chapel Hill to study the performance of the LEGEND-200 production electronics in LAr prior to the initial detector deployment. This test stand allows 1-2 HPGe detectors to be deployed and operated in LAr using the LEGEND-200 detector holders, electronics and digitizers. It also allows for quick turnaround testing and prompt feedback to hardware and electronics groups.

Figure 1 presents a schematic of the FCT. The test stand consists of a cryostat filled with LAr, an upper chamber with flanges for head electronics (HE) crates, which control and monitor the settings of the signal amplification electronics, and high voltage (HV) crates, which hold high voltage filters and interlocks, a HPGe detector mounting apparatus with an infrared (IR) shield, a radioactive source insertion tube, and a winch system with an external handle to raise/lower the detector unit.

Identical to the LEGEND-200 detector unit, the detector rests on 3 plastic insulators attached to a polyethylene naphthalate (PEN) base plate. The PEN base plate houses receptacles for the front-end electronics and HV, providing stability to establish electrical contacts on the detector via a wire bond. A trio of copper rods completes the detector unit and secures it to the IR shield that surrounds a majority of the detector. A steel band spans the distance between the winch and the IR shield allowing the detector to be raised and lowered.

An MPOD unit supplies voltage that passes through a HV filter board before entering the cryostat via connectors on the HV flange. A production HV cable bundle carries the HV signal from the flange to the detector unit, where a connector slides into one of the receptacles on the PEN base plate. The other side of the receptacle is wire bonded to the HV contact of the detector.

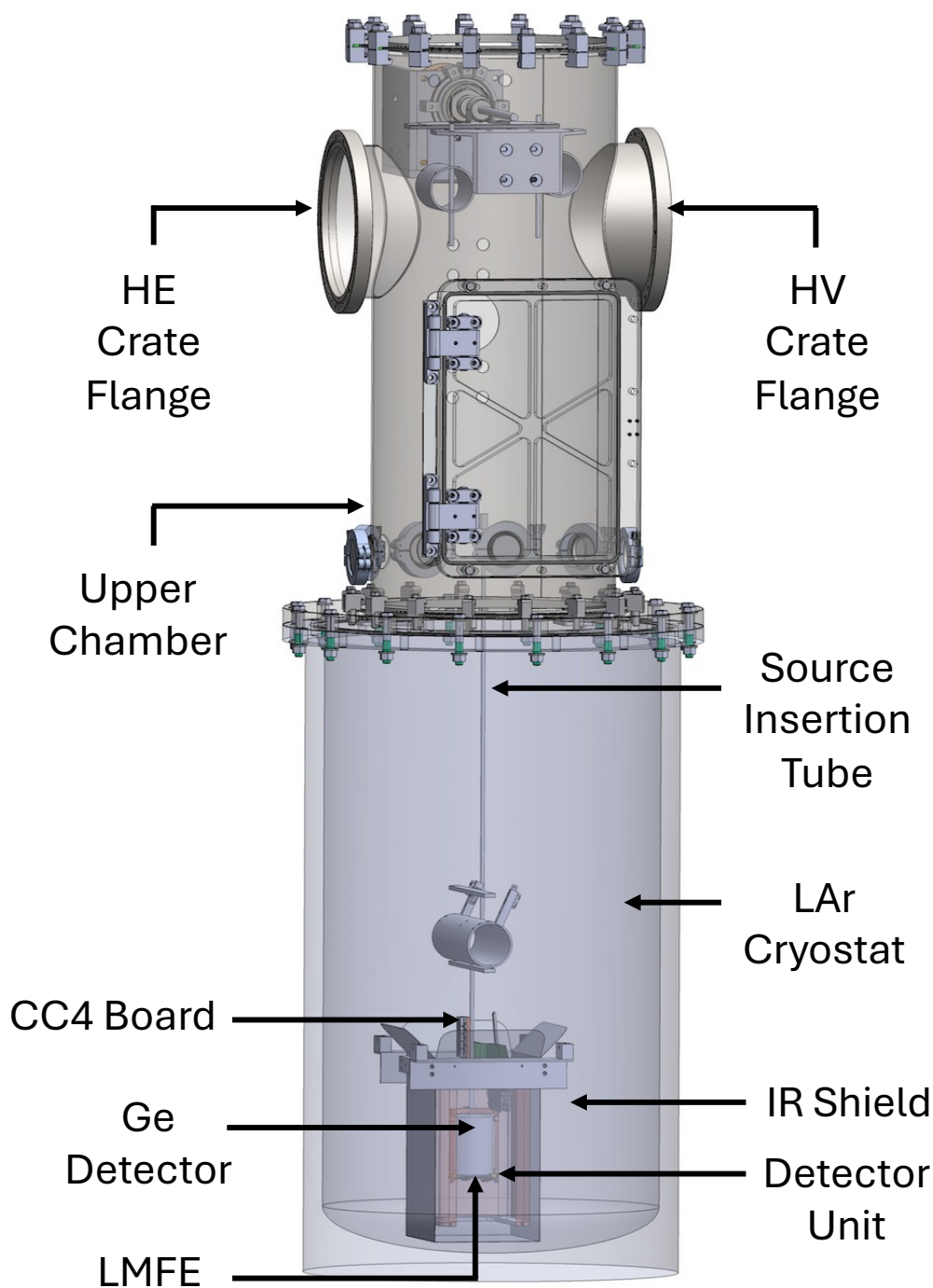

Figure 1: Full Chain Test experimental setup schematic. The stand consists of an upper chamber on top of a liquid Argon (LAr) cryostat. Inside the LAr is the detector unit holding an Inverted Coaxial Point-Contact (ICPC) detector, within an infrared (IR) shield and connected to a winch system. The Low Mass Front End (LMFE) is wire bonded to the bottom of the detector via a phosphor-bronze spring and connected to the CC4 charge-sensitive amplifying board on the top plate of the IR shield. The top plate holds connections to the head electronics (HE) and high voltage (HV) crates.

The signal extraction from the detector begins with a similar wire bond from the readout contact of the detector to the other receptacle on the PEN base plate. The first stage of signal amplification, the Low Mass Front-End (LMFE) [24], slides into this receptacle. Axon' pico-coaxial cables connect the LMFE to a 'CC4' board that is mounted on the top plate of the IR shield, which provides the second stage of amplification [25]. The CC4 board contains 7 channels; one channel is connected to the ICPC detector and the remaining 6 are connected to "dummy" boards containing capacitive loads. A long Kapton cable band transmits the signals to the HE flange for extraction from the cryostat by HE cards. These cards also control the voltage settings of the CC4 and LMFE. An external square wave pulse is injected into the FCT readout electronics to provide test signals for the ICPC detector and the dummy boards. All the electronics components utilized in the FCT are the same as those used LEGEND-200 production, including full length cables.

The extracted signals are then digitized by a FlashCam [26] analog-to-digital converter (ADC) card before being stored on a local Mac mini machine. The data readout and storage from the FCT is managed by Object-oriented Real-time Control and Acquisition (ORCA) [27] software. The sampling frequency of FlashCam is 62.5 MHz and the individual waveform trace length recorded by ORCA is 8,192 samples. The data are then decoded into HDF5 format and signals are processed in Python with the `pygama` framework [28, 29, 30, 31].

Our setup contains one FlashCam card with 6-channel inputs and outputs. We record data for 5 dummy boards and the ICPC detector. We use the dummy board data as a proxy for data taken under an environment with a low rate of physics events. Since the FCT is not shielded, the background rate of physics events is high. Thus, a radioactive source was not necessary to obtain high-rate data for the ICPC detector. For this study, we recorded data for all 6 channels for a period of 24 hours. The data collected for each system is treated separately to demonstrate our model's performance in different configurations. Thus, we curated datasets for each system to train two versions of our ML model, which we present in the next section.

## 3. Model summary

The ML-powered data cleaning mechanism we used consisted of three steps: (1) extract pulse shape information from waveforms, (2) group similar waveforms with an unsupervised learning model and re-label them based on user input, (3) extend clustering power to larger datasets with a supervised learning model. Figure 2 illustrates the process followed by the ML-powered data cleaning mechanism.

For the first step, we utilize a Discrete Wavelet Transform (DWT), which has been previously adopted for low energy analyses in Germanium-based experiments [32, 33]. The DWT decomposes the waveform into mutually orthogonal down-sampled sets of time-series coefficients by convolving the input signal with a given type of wavelet. The DWT can be performed multiple times on the same input signal, resulting in a multilevel decomposition with a down-sampling factor of $2^l$, where $l$ is the level or number of decompositions to be performed. The DWT outputs approximation coefficients (AC) and detail coefficients (DC), which capture low and high frequency components of the input signal respectively. Thus, the DWT serves to de-noise and reduce the dimension of the input waveform. In our model, we use a multilevel DWT decomposition with a value of $l$ that satisfies:

$$d = \frac{n}{2^l} \leq 256 \qquad (1)$$

where $n$ is the number of samples of the input waveform and $d$ is the number of samples of the AC. We take the AC as a lower-dimensional representation of the input waveform.

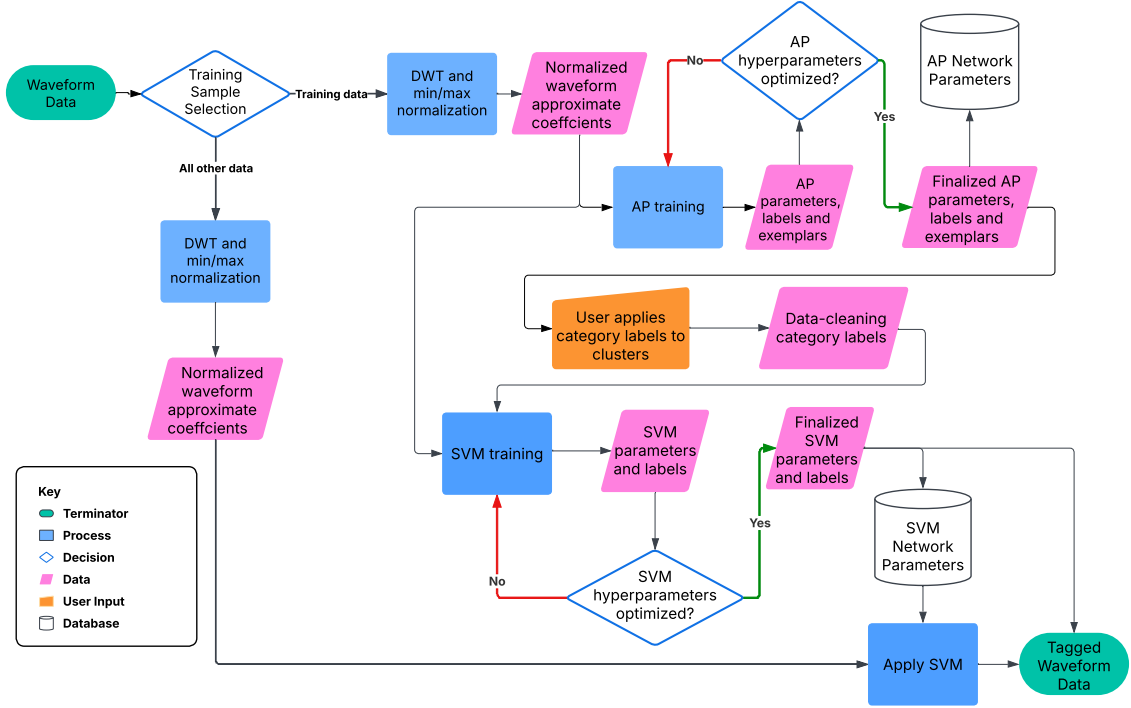

Figure 2: Schematic diagram illustrating the process adopted by the ML-powered data cleaning model.

In general, HPGe waveform data always includes electronic noise and effects from digitization of the signal, but important information about the charge drift in the detector is contained in the rising edge and sharp turnover from the rising edge to the tail. Consequently, we utilize Haar wavelets for the DWT decomposition as they preserve this information. Haar wavelets have previously been used for similar tasks in HPGe detectors, as in [34]. Since our input waveforms contain 8,192 samples, we use a value of $l = 5$ to reduce the dimension to 256 samples. We finally normalize the AC by the absolute value of their maximum or minimum amplitude, whichever one is greater, such that the values of the AC lie in the $[-1, 1]$ range. We refer to this process as *max/min* normalization.

Once we have de-noised, down-sampled, and normalized the input waveforms, we proceed to group and label them using AP. AP is an unsupervised learning algorithm that clusters inputs based on a message-passing method between data points [20]. The number of clusters is automatically computed by AP, and each cluster center is labeled as an "exemplar." The algorithm is initiated by computing the negative distances between all waveforms and storing them in the "similarities" matrix $S$. For our study, the measure of distance we utilize is the Manhattan distance, also known as L1 norm, as it captures absolute changes in discrete waveform data making it robust to small shifts. Consequently, $S$ is given by:

$$S_{ik} = -\sum_{j=1}^{n} |x_j^i - x_j^k| \tag{2}$$

$$S_{kk} = p \tag{3}$$

where $x^{i,k}$ correspond to waveforms $i$ and $k$, $n$ is the number of samples in each waveform and $p$ is a hyperparameter known as the "preference." The value of $p$ controls the number of clusters found by AP, with larger values of $p$ leading to more clusters. If p $\geq$ 0 every waveform would choose itself as an exemplar, so $p$ must be set to a negative value to produce clustering. The value of $p$ is initially set to the median of all similarities, but this hyperparameter can be tuned to produce a desired number of clusters.

After computing the similarities, the message-passing process between waveforms begins. For this stage, three additional matrices are defined: the availability $A$, responsibility $R$, and criterion $C$ matrices, with all their elements initialized to 0. Waveform $i$ sends its responsibility $R_{ik}$ to a candidate exemplar waveform $k$, which reflects how well-suited $k$ is to serve as an exemplar for $i$. Then, candidate exemplar waveform $k$ replies with its availability $A_{ik}$ to waveform $i$, which reflects how appropriate $k$ is to become an exemplar for $i$. The responsibilities are calculated as follows:

$$R_{ik}[t] = R_{ik}[t-1] \cdot \lambda + (1-\lambda) \cdot \left( S_{ik} - \max_{j \neq k, i} (S_{ij} + A_{ij}[t-1]) \right) \tag{4}$$

where $\lambda$ is a damping factor between 0.5 and 0.99 added for algorithmic convergence and $t$ is the iteration index running from $t = 1, \ldots, T$, where $T$ is the maximum number of iterations. The availabilities are computed according to:

$$A_{ik}[t] = A_{ik}[t-1] \cdot \lambda + (1-\lambda) \cdot \min\left( 0, R_{kk}[t-1] + \sum_{j \neq i, k}^{N} \max(0, R_{jk}[t-1]) \right) \tag{5}$$

$$A_{kk}[t] = A_{kk}[t-1] \cdot \lambda + (1-\lambda) \cdot \sum_{j \neq k}^{N} \max(0, R_{jk}[t-1]) \tag{6}$$

After every iteration, the criterion matrix is updated by $C_{ik} = A_{ik} + R_{ik}$. For waveform $i$, the value of $k$ that maximizes the criterion, $k_{max}$, identifies the exemplar. If $k_{\max} = i$, $i$ taken as an exemplar, and if $k_{\max} \neq i$, then waveform $k_{\max}$ is the exemplar for $i$. The message-passing process continues until all values in the criterion matrix remain

Machine Learning-Powered Data Cleaning for LEGEND

unchanged for a specified number of iterations ($\tau$), at which point AP has converged. The algorithm stops at iteration $t = T$ if this convergence condition is not met.

AP automatically computes the number of exemplars and assigns labels to all the waveforms in the training dataset. We choose AP over other algorithms that automatically identify the number of clusters in the training data, such as DBSCAN [35], as our data contains clusters of non-uniform densities. AP, however, is memory-intensive since it must store four $N \times N$ matrices until convergence, where $N$ is the number of waveforms in the training dataset. Thus, we can only train AP for datasets with $N \leq 10,000$ events. Other algorithms that can handle large training datasets efficiently, such as Balanced Iterative Reducing and Clustering using Hierarchies (BIRCH) [36], as they do not automatically compute the number of clusters.

In order to expand the labelling power of AP to datasets with $N > 10,000$ events, we utilize a SVM. The SVM is a supervised learning algorithm that classifies inputs into distinct categories by drawing hyperplanes based on "support vectors." We choose a SVM over other supervised learning classifiers as it excels at handling high-dimensional data (unlike k-Nearest Neighbors [37] or Gaussian Processes [38]) while providing a directly interpretable and intuitive explanation of how inputs are separated (unlike Random Forests [39] or Neural Networks [40]).

The SVM consists of labeled input data $(\mathbf{x}_1, y_1), ..., (\mathbf{x}_N, y_N)$ where the labels $y_i \in \{1, -1\}$. The goal of the algorithm is to find the hyperplane that maximizes the distance, or the margin, between all $y_i = 1$ inputs and all $y_i = -1$ inputs. For data that are not linearly separable, it is necessary to employ the kernel trick [41]. The kernel trick consists of transforming the input data samples $\mathbf{x}_i$ into another dot product feature space via:

$$K(\mathbf{x}_i, \mathbf{x}_j) = \langle \phi(\mathbf{x}_i), \phi(\mathbf{x}_j) \rangle \qquad (7)$$

where $\phi(\mathbf{x})$ is the mapping function. In our model we utilize Gaussian radial basis function (RBF) kernel as it produces smooth decision boundaries, which helps in generalizing better to unseen data. The RBF kernel is given by:

$$K(\mathbf{x}_i, \mathbf{x}_j) = \exp\left(-\gamma \|\mathbf{x}_i - \mathbf{x}_j\|^2\right) \qquad (8)$$

where $\gamma > 0$ is a modifiable hyperparameter. The objective of the SVM is to maximize:

$$\max_{\alpha_i} \left( \sum_{i=1}^{N} \alpha_i - \frac{1}{2} \sum_{i=1}^{N} \sum_{j=1}^{N} \alpha_i \alpha_j y_i y_j K(\mathbf{x}_i, \mathbf{x}_j) \right) \qquad (9)$$

subject to the constraint:

$$\sum_{i=1}^{N} \alpha_i y_i = 0, \quad 0 \leq \alpha_i \leq C, \quad \forall i \qquad (10)$$

where $\alpha_i$ are Lagrange multipliers and $C$ is a regularization parameter. Since our data contains $\kappa > 2$ labels, we utilize a multiclass SVM [42]. The multiclass formulation

Machine Learning-Powered Data Cleaning for LEGEND9trains a binary SVM for each class $\kappa$. For class $\kappa$ and input sample $i$, the decision function is given by:

$$f_\kappa(\mathbf{x}_i) = \sum_{l=1}^{N_\kappa^*} \alpha_{l,\kappa} y_{l,\kappa} K(\mathbf{x}_l, \mathbf{x}_i) + b_\kappa \qquad (11)$$

where $b_\kappa$ is the intercept of the separating hyperplane and $N_\kappa^*$ is the number of support vectors for class $\kappa$. The support vectors are the input samples that lie closest to the separating hyperplane found by each binary SVM. The multiclass SVM then assigns a label to input sample $i$ via:

$$\hat{y}_i = \arg\max_\kappa \left( f_\kappa(\mathbf{x}_i) \right) \qquad (12)$$

where $f_\kappa(\mathbf{x})$ is the decision function for class $\kappa$.

## 4. Model training and optimization

We create two separate models for each system, one for detector data and another for dummy board data. The systems contain different types of waveforms due to their operational characteristics. The high background physics event rate of our experimental setup leads to pileup in the waveform traces of the HPGe detector. This scenario is akin to having a calibration source deployed near the LEGEND-200 detectors. The dummy boards contain waveforms caused by pulsed signals, discharges and crosstalk. This resembles the data captured by HPGe detectors in a low background environment without any calibration sources in the vicinity. Thus, treating each system separately allows us to simulate our model's performance in a low background setting during calibrations and $0\nu\beta\beta$ data taking.

We train our models on a datasets containing 10,000 waveforms for each system. We first normalize the waveforms using the *max/min* method. Then, we compute the pairwise distances between normalized waveforms and store them in the similarities matrix $S$. The similarities are then fed into AP. We perform a grid search to optimize the hyperparameters of AP, namely, the preference $p$ and the damping factor $\lambda$. For each model, we search over 100 grid points spanning $\lambda \in [0.85, 0.99]$ and $p \in [\min(S), -100]$. We constrain the lower limit of $\lambda$ to 0.85 since lower values cause create algorithmic non-convergence of AP in our data. Every iteration of AP utilizes $\sim$ 12 GB of RAM, which requires the optimization process to run on multiple CPU cores in parallel. The hyperparameter combination that gives the closest to 100 clusters is then used. Figure 3 shows the hyperparameter grid used to optimize AP. The point enclosed by a red box was the combination of $p$ and $\lambda$ that gave the closest to 100 clusters, and was used to train AP.

All waveforms in the training dataset are labeled according to the cluster they chose. The waveform located at each cluster center is defined as the exemplar for that cluster. AP may find multiple clusters for a given human-defined data cleaning category. Therefore, the user manually maps the labels provided by AP for each cluster to a set of standard data cleaning tags. At this stage, a new tag can be added if the user identifies an AP cluster as representing a distinct class of waveforms. Otherwise, the user re-labels the each AP exemplar according to a data cleaning tag of the standard set that most closely resembles its shape. Figure 4 presents a comprehensive set of data cleaning tags with sample waveform plots. The re-labeled exemplars are presented in Figure 5a and 5b, where each color corresponds to a different data cleaning tag. The exemplars of Figure 5b are dominated by the Noise Trigger category due to the sparse mode configuration of the DAQ. Every time a single channel triggers, the DAQ records the waveform traces from all channels. Since the detector channel has a high rate events, most of the recorded waveforms from dummy channels are empty traces.

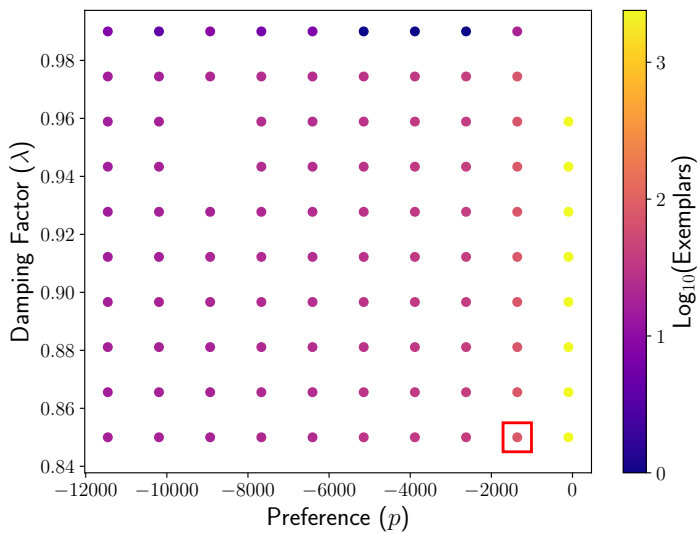

Figure 3: AP hyperparameter grid. The combination of $p$ and $\lambda$ that gave the closest to 100 exemplars, enclosed in a red box, was used to train the optimal AP. Empty points in the grid represent combinations of hyperparameters where AP did not converge.

Once the waveforms are re-labeled according to the data cleaning tags of Figure 4, we train the SVM. Since each waveform contains 8,192 samples, we first pre-process them using a DWT with 5 levels, giving AC of 256 samples according to Eq. (1). This downsampling allows the SVM to both be trained and perform predictions more quickly without compromising its accuracy. The AC are *max/min* normalized before being passed to the SVM. We optimize the hyperparameters of the SVM, $C$ and $\gamma$, with a random grid search over a broad range of values spanning several orders of magnitude: $[10^{-2}, 10^{10}]$ for $C$ and $[10^{-9}, 10^{3}]$ for $\gamma$ to ensure proper coverage of the hyperparameter space. We employ 5-fold cross validation for the optimization process, splitting the dataset into a training and validation set with an 80:20 ratio respectively. We use the cross-validated accuracy as the figure-of-merit for the optimization process given by:

$$\text{Accuracy} = \frac{1}{CV \cdot N_V} \sum_{i=1}^{CV} \sum_{j=1}^{N_V} \frac{|Y_{i,j} \cap \hat{Y}_{i,j}|}{|Y_{i,j} \cup \hat{Y}_{i,j}|} \qquad (13)$$

where $CV$ is the number of cross-validation splits, $N_V$ is the number of AC in the validation set, $Y_{i,j}$ is the set of true labels as given by AP for the $i$-th cross-validation split and the $j$-th input AC, and $\hat{Y}_{i,j}$ is the set of predicted SVM labels for the $i$-th cross-validation split and the $j$-th input AC. Furthermore, the values of $C$ were scaled inversely with the number of observations for each category according to Eq.

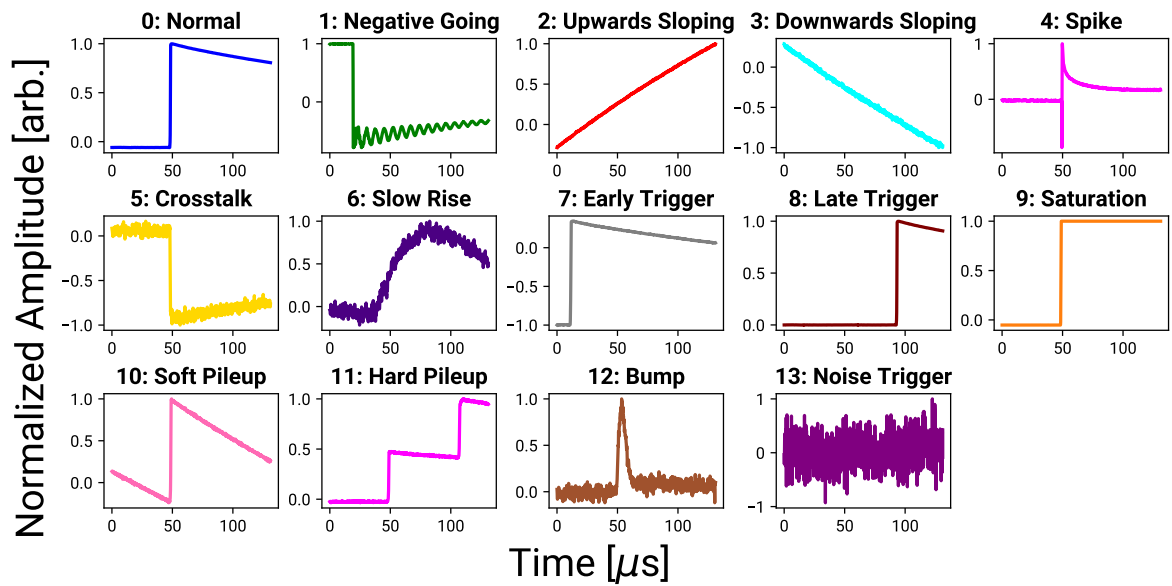

Figure 4: Data cleaning categories.

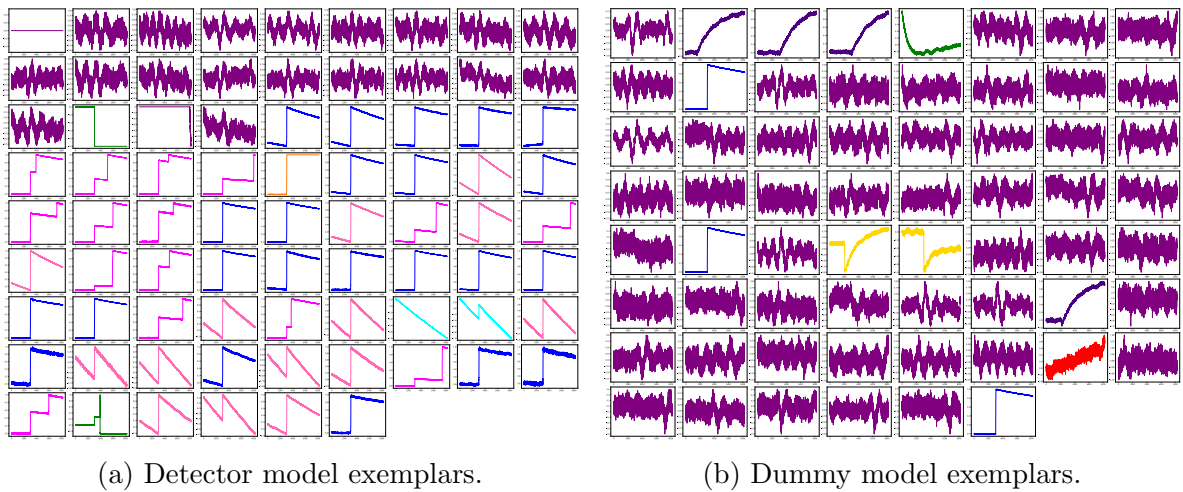

(a) Detector model exemplars.            (b) Dummy model exemplars.

Figure 5: Exemplars re-labeled according to the data cleaning categories of Figure 4.

$$C_\kappa = C \times \frac{N}{m \times N_\kappa} \tag{14}$$

where $C_\kappa$ is the scaled value of $C$ for category $\kappa$, $N$ is the number of AC in the training set, $m$ is the number of categories, and $N_\kappa$ is the number of AC pertaining to category $\kappa$. Since the Noise Trigger and Normal category dominate the dummy and detector training data respectively, we add this inverse weight scaling to avoid biasing the SVM. Hyperparameter values are randomly sampled from a log-uniform distribution to compose 500 combinations of $C$ and $\gamma$. The combination that gives the accuracy closest to 1 is then used to train the optimal SVM. Figure 6 shows the hyperparameter grid used to optimize the SVM. The point enclosed by a red box was the combination of $C$ and $\gamma$ that gave the cross-validated accuracy closest to 1, and was used to train the SVM.

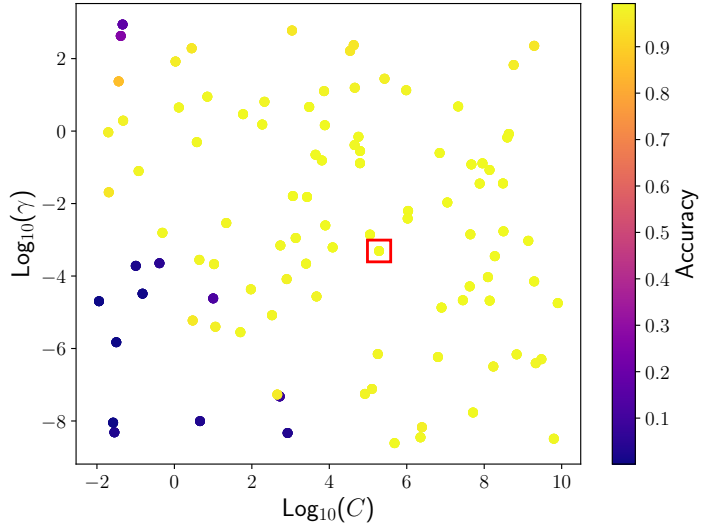

Figure 6: SVM hyperparameter grid. The combination of $C$ and $\gamma$ that gave the cross-validated accuracy closest to 1, enclosed by a red box, was used to train the optimal SVM.

To visualize the decision boundaries of the trained SVM, we must perform dimensionality reduction since the models are trained using 256-dimensional data. For this, we use a t-distributed Stochastic Neighbor Embedding (t-SNE) [43] algorithm to reduce both the AC and the SVM decision boundaries into 3D space. The 3D representations of the AC and the SVM decision boundaries are shown in Figure 7a and 7b. With the trained SVM, we can predict labels for larger datasets ($N > 10,000$).

We run the training and optimization of our model at the Longleaf computing cluster [44] of the University of North Carolina at Chapel Hill using `scikit-learn` [45, 46]. For AP, we allocate 12 GB of RAM on 6 CPU cores to run each grid point iteration of the hyperparameter search. We run the search iterations in parallel for wall time efficiency. For the SVM, we utilize the `RandomizedSearchCV` class to conduct the hyperparameter search, allocating 64 GB of RAM over 32 CPU cores running in parallel. The training and optimization of the model consumes 18 CPU hours in total.

## 5. Model performance and efficiency

To understand the performance of our model, we apply ML data cleaning cuts to the full detector and dummy datasets. Waveforms from the Normal (0) category encompass events caused by energy depositions in HPGe detectors from alpha, beta, and gamma particles. Waveforms from the Saturation (9) category are caused by highly energetic atmospheric muons that deposit energies in the HPGe detector larger than the 8 MeV dynamic range of the DAQ system. Muon events are tagged at later stages in the analysis chain via a muon veto system [47], which is not implemented in the FCT. Since we aim to keep only events caused by non-pileup physics interactions, we define the ML data cleaning cut according to Eq. 15.

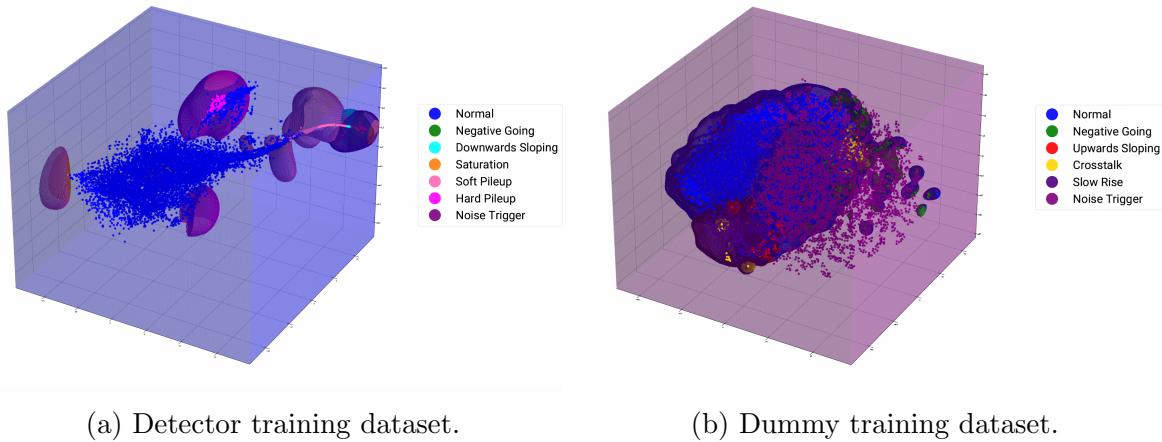

(a) Detector training dataset.    (b) Dummy training dataset.

Figure 7: 3D representations of the training datasets with their SVM decision boundaries.

$$\text{MLDataCleaningCut} = \text{SVMPrediction} \in \{0, 9\} \quad (15)$$

Waveforms pertaining to the all other categories of Figure 4 are thus rejected my the ML data cleaning cut of Eq. 15.

Figures 8 and 9 show the energy spectra of the detector and dummy datasets, respectively. The detector dataset contains $N = 4,644,992$ events, while the dummy dataset contains $N = 23,224,967$ events. Filled brown spectra correspond to all events before cuts, green spectra correspond to events accepted by the ML data cleaning cut, and magenta spectra correspond to events rejected by the ML data cleaning cut.

The main purpose of our model is to remove all anomalous events while keeping all physical events in our datasets. Thus, we want to evaluate how effective our model in terms of physical events that are incorrectly tagged as non-physical, defined as the model sacrifice, and anomalous events that are accepted, defined as the model leakage. A high leakage of anomalous events is preferred to a high sacrifice of physics events. Anomalous events that leak into our datasets are typically eliminated by pulse shape discrimination (PSD) cuts at later stages of the analysis chain [48, 49]

To evaluate the sacrifice and leakage of the model, we construct datasets that only contain waveforms of a given type using traditional data cleaning methods. Traditional data cleaning methods rely on digital signal processing parameters that are calculated directly from waveforms. Table 1 summarizes the traditional data cleaning parameters used to isolate different categories of waveforms for sacrifice and leakage studies.

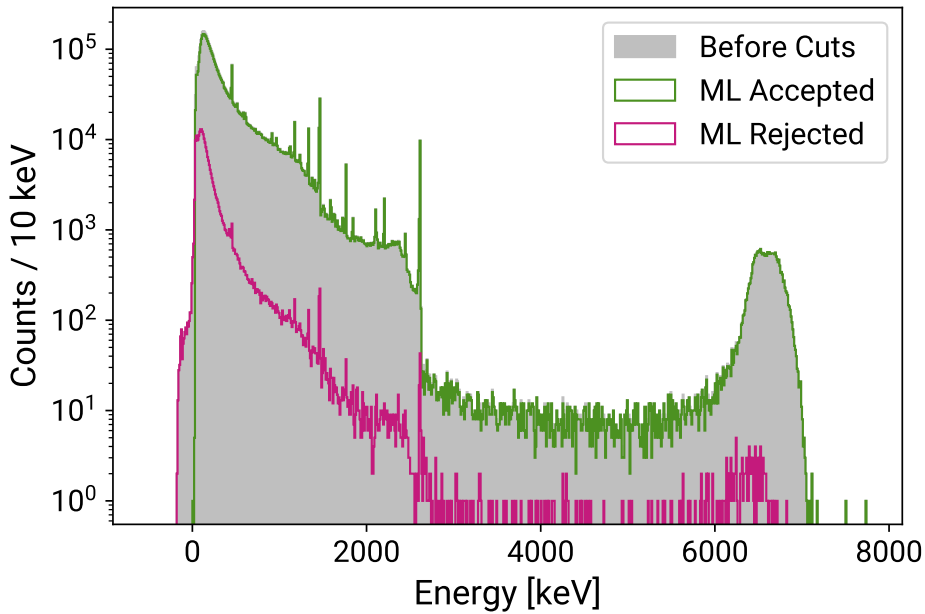

Figure 8: Detector energy spectra before and after applying ML data cleaning cuts. The spectrum before cuts, shown in gray, is characterized by gamma peaks from the $^{238}$U and $^{232}$Th chains and energies below 0 keV corresponding to negative-going waveforms caused by discharges. The broad peak centered around 6.5 MeV corresponds to muon events that saturate the dynamic range of the DAQ system. The spectra of accepted and rejected events by the ML data cleaning cut of Eq. 15 are shown in green and magenta, respectively.

### 5.1. Sacrifice

To perform sacrifice studies, we construct datasets of physics events. We pre-apply data cleaning cuts based on traditional parameters to ensure the sacrifice datasets contain only waveforms caused by physics interactions in the HPGe detector. To get an estimate of the physics event sacrifice of our models, all waveforms of the datasets are assigned data cleaning labels using the trained SVMs. We then apply the ML data cleaning cut defined in Eq. (15). The physics event sacrifice $s$ for a given dataset is defined as the ratio of rejected events $N_r$ to total events $N$ as per Eq. (16). The uncertainties on the event sacrifice are statistical and calculated using 90% Clopper-Pearson confidence intervals [50].

$$s = \frac{N_r}{N} \qquad (16)$$

Table 2 presents the sacrifice estimates for the physics event categories per model. The detector model presents a sacrifice of $0.024^{+0.004}_{-0.003}\%$ on the Normal category. The energies of all rejected events lie below 150 keV as shown in Figure 10. In this energy region the signal-to-noise ratio is reduced, and it becomes more difficult for the model to disentangle physics signals from anomalous populations. Figure 11 depicts sample rejected waveforms by the detector model in the Normal category sacrifice dataset. These low energy waveforms are characterized by a slow charge collection component at the top of the rising edge and on the tail. Dedicated studies have shown that this slow charge collection is associated with interactions occurring near the detector surface, a source of backgrounds for $0\nu\beta\beta$ searches [51, 52]. Such events are targeted for removal by subsequent stages of the analysis, so they represent a sacrifice for the data cleaning stage. The dummy model yields a 0% sacrifice in the Normal category, which is expected as all the waveforms are generated from a test pulse injector at energies over 400 keV.

Table 1: Traditional data cleaning parameters.

| Parameter Name | Description |
| --- | --- |
| Waveform Slope | Slope of the full waveform. |
| Baseline Mean | Average of the first 45 $\mu s$ of the waveform. |
| Baseline Slope | Slope of the first 45 $\mu s$ of the waveform. |
| Baseline St. Dev. | Standard deviation of the first 45 $\mu s$ of the waveform. |
| Tail St. Dev. | Standard deviation of the last 20 $\mu s$ of the waveform following pole-zero correction. |
| Trapezoidal Max (Min) | Maximum (minimum) amplitude of the trapezoidal-filtered waveform. |
| Time Point $X$ | Time point corresponding to $X$ percentage of the waveform's maximum amplitude. |
| High (Low) Saturation Time Point | Time point corresponding to the first instance of the waveform reaching the high (low) ADC saturation value. |
| Inverted Time Point $X$ | Time point corresponding to $X$ percentage of the inverted waveform's maximum amplitude. |
| Inverted Trapezoidal Max (Min) | Maximum (minimum) amplitude of a trapezoidal filter on the inverted waveform. |
| Baseline Pileup Max (Min) | Maximum (minimum) amplitude of the first 45 $\mu s$ of the trapezoidal-filtered waveform using a short integration time. |
| Tail Pileup Max (Min) | Maximum (minimum) amplitude of the last 65 $\mu s$ of the trapezoidal-filtered waveform using a short integration time. |
| Trapezoidal Fixed-Time Pickoff | Value of the trapezoidal-filtered waveform at 17.6 $\mu s$ after the start of the waveform rise. |
| Effective Drift Time | Time taken for the waveform to reach its maximum amplitude corrected for charge-trapping and multi-site event effects. |
| Energy | Estimate of the event's energy from a trapezoidal filter with optimized parameters after ADC to keV calibration. |

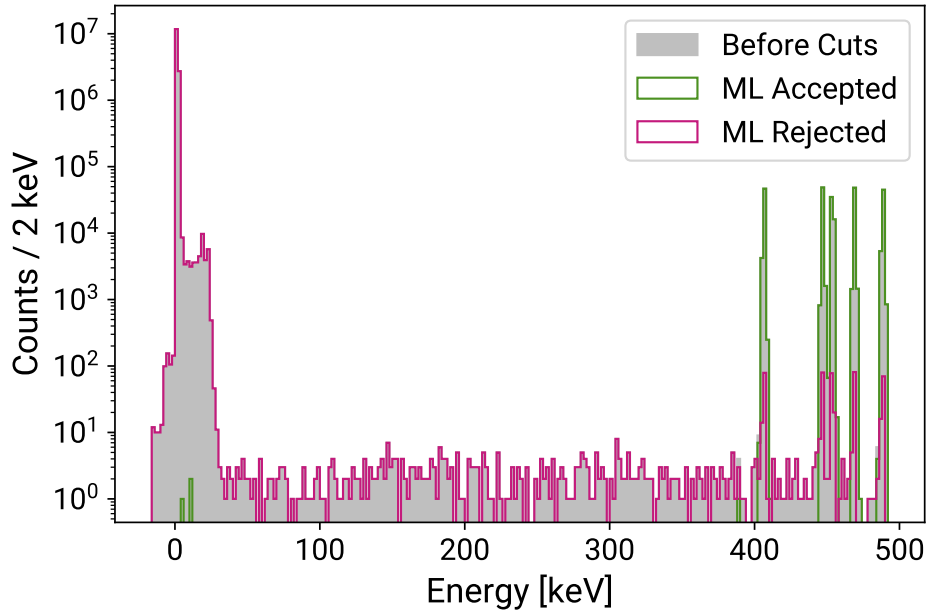

Figure 9: Dummy energy spectra before and after ML data cleaning cuts. The spectrum before cuts, shown in gray, is characterized by 5 peaks above 400 keV corresponding to injected test pulses for each dummy board. The peak at 0 keV corresponds to noise triggers recorded by the global trigger settings of the DAQ system. The events below 0 keV correspond to discharges and negative polarity crosstalk waveforms from the injected test pulses. The events in between 0 and 50 keV correspond to positive polarity crosstalk waveforms from muon events in the detector channel, and upwards sloping waveforms caused by recovery to baseline from previous discharges. The spectra of accepted and rejected events by the ML data cleaning cut of Eq. 15 are shown in green and magenta, respectively.

Table 2: Physics event sacrifice of ML data cleaning cuts. Estimates are included only for categories found by AP during training.

| Category | Detector Model | | Dummy Model | |
| --- | --- | --- | --- | --- |
| | $N$ | $s$ (%) | $N$ | $s$ (%) |
| Normal | 541,952 | $0.024^{+0.004}_{-0.003}$ | 14,603 | $0.000^{+0.021}_{-0.000}$ |
| Saturation | 23,659 | $0.000^{+0.013}_{-0.000}$ | - | - |

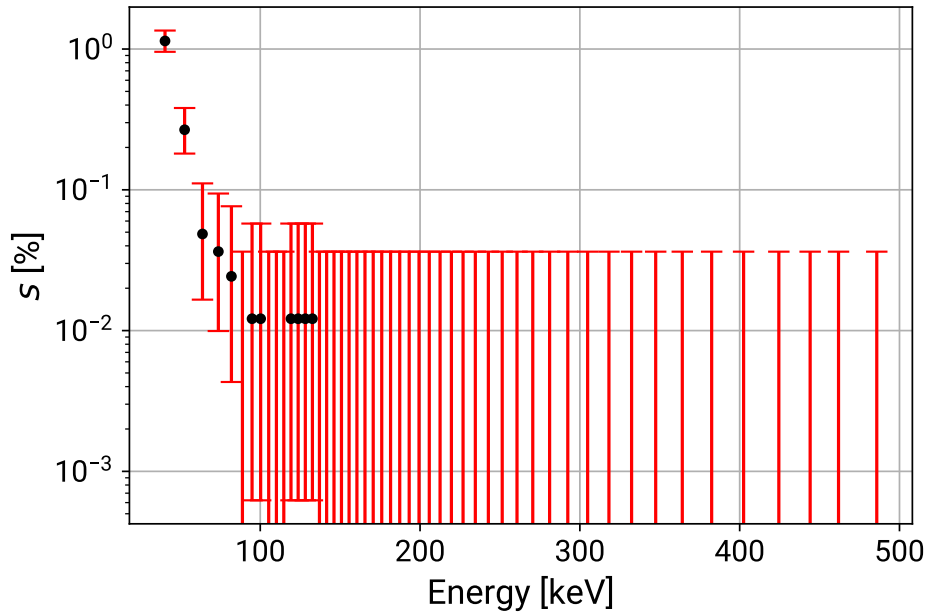

Figure 10: Energy dependence of the sacrifice in the Normal category detector dataset. Variable quantile binning is used. For events with energies above 150 keV, the sacrifice is minimal at 0 %, thus only upper boundaries on the statistical uncertainty are shown.

The extremely low sacrifice, particularly at energies over 150 keV, make the AP-SVM model highly suitable for experiments searching for $0\nu\beta\beta$, which would occur in a narrow Region-of-Interest around the $^{76}$Ge $\beta\beta$ decay Q-value of 2039 keV. Past $0\nu\beta\beta$ experiments using HPGe detectors have achieved data cleaning sacrifice levels below 0.1% [3, 4], which the AP-SVM method is able to match or improve upon. The resulting data-cleaning signal efficiency is combined with other analysis cut efficiencies to determine the overall signal acceptance of an experiment; the future LEGEND-1000 experiment is targeting an overall analysis cut efficiency of at least 90%. The sacrifice level demonstrated for AP-SVM makes it an appropriate method for this future experiment.

*5.2. Leakage*

To perform leakage studies, we construct datasets of anomalous events. We pre-apply data cleaning cuts based on traditional parameters to ensure the leakage datasets contain only waveforms caused by non-physical interactions. To get an estimate of the anomalous event leakage of our models, all waveforms of the datasets are assigned data cleaning labels using the trained SVMs. We then apply the ML data cleaning cut defined in Eq. (15). The anomalous event leakage $l$ for a given dataset is defined as the ratio of accepted events $N_a$ to total events $N$ as per Eq. (17). Again, the uncertainties on the event leakage are statistical and calculated using 90% Clopper-Pearson intervals.

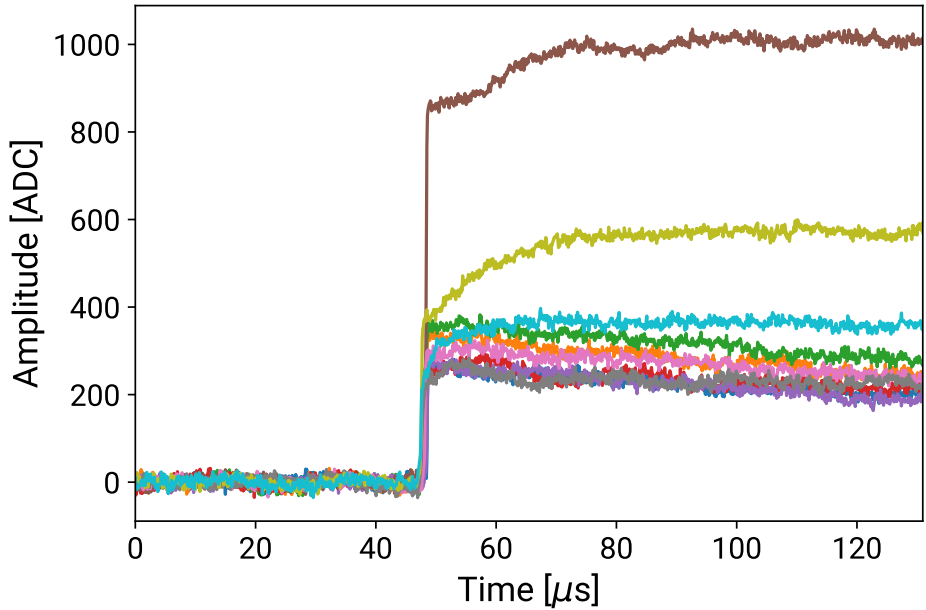

Figure 11: Sample rejected waveforms from the Normal category detector sacrifice dataset. Rejected waveforms are characterized by low-frequency noise and slow charge collection in the rising edge and tail.

$$l = \frac{N_a}{N} \qquad (17)$$

Table 3 presents the leakage estimates for the anomalous event categories per model. The Upwards Sloping and Crosstalk categories are caused by recoveries from discharges to baseline and induced signals from neighboring channels, respectively. These are absent from the detector model due to the high rate of physics events, which dominate the triggering scheme of the DAQ system. The Downwards Sloping category is caused by recoveries to baseline after saturation of the DAQ system, which are not present in the dummy model since no muon events are captured by the dummy boards. Anomalous event leakage is seen in the detector model for the Soft Pileup, Hard Pileup, and Noise Trigger categories. The considerable leakage of 13 and 15% on the Soft and Hard Pileup categories can be attributed to the resemblance of the waveforms to those in the Normal category. The downwards sloping baseline of Soft Pileup waveforms is caused by the decaying tail of a previous waveform, as shown in Figure 12a. The very slight negative slope on the baseline is below the tagging threshold defined by the SVM, causing it to classify these waveforms as Normal. The accepted Hard Pileup waveforms are characterized by a considerably lower energy event riding on top of the decaying tail of the first event in the same waveform trace, as shown in Figure 12b. The small magnitude of the second event compared to the first causes the SVM to classify these waveforms as Normal. In general, it is not possible to define a clear separation boundary between Normal and Soft/Hard Pileup waveforms in this data, given the presence of low frequency noise and multi-site events. All data cleaning methods must choose an acceptable level of leakage for these categories. In building the AP-SVM model, we prefer to keep the physics event sacrifice low while incurring a moderate leakage of pileup waveforms, leading to more conservative data cleaning.

Table 3: Anomalous event leakage of ML data cleaning cuts. Estimates are included only for categories found by AP during training.

| Category | Detector Model | | Dummy Model | |
| --- | --- | --- | --- | --- |
| | $N$ | $l$ (%) | $N$ | $l$ (%) |
| Negative Going | 319 | $0.000^{+0.935}_{-0.000}$ | 907 | $0.000^{+0.330}_{-0.000}$ |
| Upwards Sloping | - | - | 2,637 | $0.000^{+0.114}_{-0.000}$ |
| Downwards Sloping | 1,151 | $0.000^{+0.260}_{-0.000}$ | - | - |
| Crosstalk | - | - | 5,410 | $0.000^{+0.055}_{-0.000}$ |
| Slow Rise | - | - | 3,247 | $0.000^{+0.092}_{-0.000}$ |
| Soft Pileup | 24,531 | $13.212^{+0.361}_{-0.354}$ | - | - |
| Hard Pileup | 69,684 | $15.034^{+0.224}_{-0.222}$ | - | - |
| Noise Trigger | 203 | $0.985^{+2.084}_{-0.810}$ | 14,370,779 | $0.000^{+0.000}_{-0.000}$ |

In the context of $0\nu\beta\beta$ searches, this conservative approach is appropriate, as the HPGe detectors operate in an extremely low-rate environment, with typical per-detector physics event rates well below 0.01 Hz during low-background data-taking. Therefore, true Pileup events are extremely rare. Pileup tags are implemented in data cleaning to facilitate energy and pulse-shape parameter calibration using higher-rate radioactive source calibration data, but providing pristine data sets for these purposes is of lower priority than maintaining high $0\nu\beta\beta$ signal efficiency. When AP-SVM is applied to LEGEND-200 data, it is paired with a traditional data cleaning tag to address Soft Pileup events, which can be set more liberally or conservatively depending on the analysis being conducted. The small fraction of surviving hard Pileup events typically does not bias parameter calibrations, and these events are allowed to remain untagged.

The two Noise Trigger waveforms that are classified as Normal, shown in Figure 13, can be attributed to the low statistics of Noise Trigger samples in the detector training dataset and the large amplitude of low frequency noise in this waveforms. The dummy training dataset is dominated by Noise Trigger samples, leading to a leakage of 0% in this category with several orders of magnitude of more samples in the leakage dataset. In fact, the dummy model demonstrates 0% leakage in all categories. This indicates that our model presents a very accurate separation between anomalous categories in our proxy for low background data.

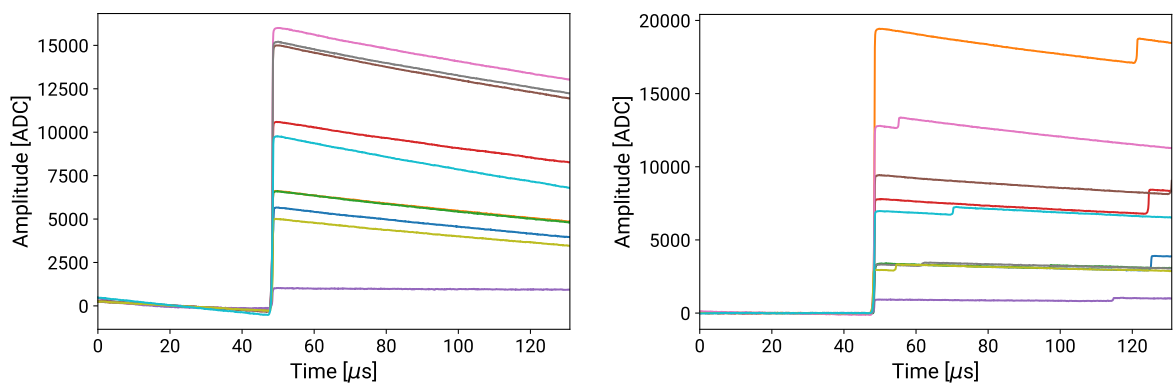

(a) Sample accepted waveforms from the Soft Pileup category detector leakage dataset.

(b) Sample accepted waveforms from the Hard Pileup category detector leakage dataset.

Figure 12: Sample accepted waveforms from the Soft and Hard Pileup category detector leakage datasets.

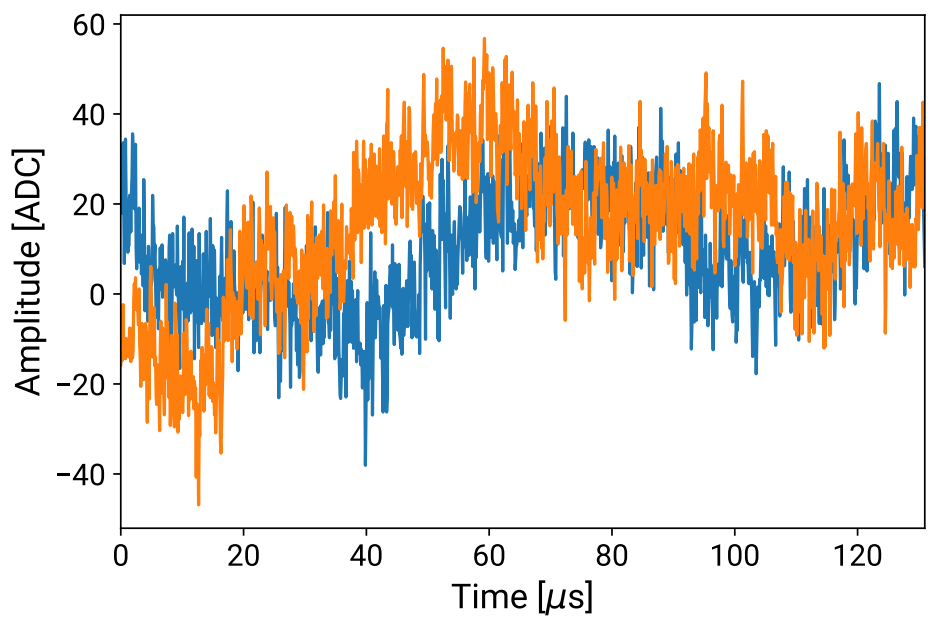

Figure 13: Accepted waveforms from the Noise Trigger category detector leakage dataset.

## 6. Conclusions

In this study, we have presented a machine learning-powered data cleaning mechanism for the LEGEND experiment. Our model is based on a clustering (AP) plus human supervision and classification (SVM) scheme to distinguish between signals produced in HPGe detectors. Utilizing data from the FCT test stand at UNC-CH, we have demonstrated that our mechanism efficiently separates signals caused by physics events from signals created by transient anomalies.

Our model presents a maximum physics event sacrifice of $0.024^{+0.004}_{-0.003}\%$ of physics signals in high rate data, where all rejected events lie below 150 keV and likely originate from the HPGe detector's surface, and $0.000^{+0.021}_{-0.000}\%$ in low rate data. Similarly, the model yields a maximum anomalous event leakage of $15.034^{+0.224}_{-0.222}\%$ in high rate data caused primarily by pileup events, and $0.000^{+0.330}_{-0.000}$ in low rate data. This performance matches or exceeds that of traditional data cleaning methods used in past HPGe-based $0\nu\beta\beta$ experiments [3, 4].

In addition to the FCT data presented in this study, we have applied our model to data from several experimental setups with different configurations. We have successfully demonstrated this data cleaning mechanism in an initial test deployment for LEGEND-200 prior to infrastructure upgrades (the post-GERDA test), the initial commissioning deployment with 60 kg of HPGe detectors, the 2023-2024 deployment with 142 kg of HPGe detectors, and in vacuum cryostat HPGe characterization stands at Oak Ridge National Laboratory and the University of North Carolina at Chapel Hill. These data sets differed in overall noise levels. The primary effect on the AP-SVM technique of this variation is that a larger variety of transient noise sources leads to more Noise Trigger category clusters being identified at the AP training stage. Since these categories are merged by the user for the SVM stage, this does not affect the overall AP-SVM model performance.

The AP-SVM technique can also be used, with minimal adaptation, for other sources of time-series data. We have tested the model performance, with success, on signals obtained from silicon photo multiplier (SiPM) detectors in LEGEND-200 [53]. The different data-acquisition mode used for LEGEND SiPM detectors, and the different waveform shape of these signals, requires a differing event pre-processing approach. Since the LEGEND-200 data acquisition system is set to trigger on HPGe signals, with SiPM waveforms captured in coincidence, photoelectron signals are not centered in the digitization window, and several signals may appear within a single windowed trace. Instead of down-sampling these waveforms using a DWT, we use an offline trigger to window each possible signal, then perform min/max normalization. These windowed signals replace the normalized approximate coefficients in the network training and application workflow shown in Fig. 2. This approach is showing promise, particularly in identifying cross-talk events that other data cleaning approaches fail to tag efficiently. A future publication is planned demonstrating this application.

Another potential application is in monitoring of environmental controls such as temperatures, detector leakage currents, vacuum pressures, and the like, where the model can be used to categorize behaviors and identify unusual excursions. The versatility of our model allows it to be utilized in experiments with time-series waveform data, quasi-agnostic to the underlying detector system, as we have demonstrated in this analysis.

With regard to data cleaning for LEGEND-1000, the AP-SVM model presents a scalable solution. Although the size of the training dataset is limited to $N \leq 10,000$ waveforms due to the memory demands of AP, evaluating the trained SVM on out-of-sample data scales only with the number of features in each waveform. The number of features, given by the dimension of the AC as per Eq. 1, remains constant. Thus, applying the SVM to larger datasets does not present a challenge for scalability of the model. Nevertheless, to ensure that the training dataset is representative of all possible categories, a uniform sampling of events throughout the entire energy spectrum and across detectors will have to be performed. Thus, AP-SVM can be incorporated in LEGEND-1000 solely by modifying the waveform selection process for training with no changes to the model itself.

Performing data cleaning procedures with traditional parameters requires significant time and human effort for even a single detector and can require frequent modification, particularly as run conditions change. AP-SVM allows for significant labor savings compared to the traditional method. Additionally, it is being used to supplement and improve upon traditional data cleaning procedures in LEGEND. For example, AP-SVM has been used to identify new anomalous populations and to study event leakage in the traditional method. The model is also being used as the main data cleaning method in LEGEND's Julia-based secondary software stack: `Juleana` [54]. The AP-SVM model can thus be utilized for data cleaning on its own, to cross-validate traditional methods, or in conjunction with traditional procedures to provide a robust data cleaning method for LEGEND.

**Acknowledgments**

This material is based upon work supported by the U.S. Department of Energy, Office of Science, Office of Nuclear Physics, under Award Numbers DE-SC0022339, DE-FG02-97ER41041, and DE-FG02-97ER41033. We acknowledge support from the Nuclear Physics Program of the National Science Foundation through grant number PHY-1812374. This work is done in support of the LEGEND-200 experiment and we thank our collaborators for their input. We would like to thank the University of North Carolina at Chapel Hill and the Research Computing group for providing computational resources and support that have contributed to the results presented in this study.

**References**

[1] N. Abgrall, I. Abt, M. Agostini, et al. LEGEND-1000 Preconceptual Design Report, 2021.

[2] Matteo Agostini, Giovanni Benato, Jason A. Detwiler, Javier Menéndez, and Francesco Vissani. Toward the discovery of matter creation with neutrinoless $\beta\beta$ decay. *Rev. Mod. Phys.*, 95:025002, May 2023.

[3] M. Agostini, G. R. Araujo, A. M. Bakalyarov, et al. Final Results of GERDA on the Search for Neutrinoless Double-$\beta$ Decay. *Phys. Rev. Lett.*, 125:252502, Dec 2020.

[4] I. J. Arnquist, F. T. Avignone, A. S. Barabash, et al. Final Result of the MAJORANA DEMONSTRATOR's Search for Neutrinoless Double-$\beta$ Decay in $^{76}$Ge. *Phys. Rev. Lett.*, 130:062501, Feb 2023.

[5] S. Mertens, N. Abgrall, F. T. Avignone, et al. MAJORANA Collaboration's Experience with Germanium Detectors. *Journal of Physics: Conference Series*, 606(1):012005, apr 2015.

[6] M. Agostini, M. Allardt, E. Andreotti, et al. Production, characterization and operation of $^{76}$Ge enriched BEGe detectors in GERDA. *The European Physical Journal C*, 75(2):39, 2015.

[7] R.J. Cooper, D.C. Radford, P.A. Hausladen, and K. Lagergren. A novel HPGe detector for gamma-ray tracking and imaging. *Nuclear Instruments and Methods in Physics Research Section A: Accelerators, Spectrometers, Detectors and Associated Equipment*, 665:25–32, 2011.

[8] M. Agostini, G. Araujo, A. M. Bakalyarov, et al. Characterization of inverted coaxial $^{76}$ge detectors in GERDA for future double-$\beta$ decay experiments. *The European Physical Journal C*, 81(6), June 2021.

[9] Erica Andreotti. Characterization of BEGe detectors in the HADES underground laboratory. *Nuclear Instruments and Methods in Physics Research Section A: Accelerators, Spectrometers, Detectors and Associated Equipment*, 718:475–477, 2013.

[10] Iris Abt, Chris Gooch, Felix Hagemann, et al. A novel wide-angle Compton Scanner setup to study bulk events in germanium detectors. *The European Physical Journal C*, 82(10), October 2022.

[11] M Agostini, L Pandola, and P Zavarise. Off-line data processing and analysis for the GERDA experiment. *Journal of Physics: Conference Series*, 368(1):012047, jun 2012.

[12] J. Myslik, N. Abgrall, S. I. Alvis, et al. Data quality assurance for the majorana demonstrator, 2017.

[13] P. Holl, L. Hauertmann, B. Majorovits, O. Schulz, M. Schuster, and A. J. Zsigmond. Deep learning based pulse shape discrimination for germanium detectors. *The European Physical Journal C*, 79(6):450, May 2019.

[14] I. J. Arnquist, F. T. Avignone, A. S. Barabash, et al. Interpretable Boosted-Decision-Tree Analysis for the MAJORANA DEMONSTRATOR. *Phys. Rev. C*, 107:014321, Jan 2023.

[15] Zhenghao Fu, Christopher Grant, Dominika M. Krawiec, Aobo Li, and Lindley A. Winslow. Generative models for simulation of KamLAND-Zen. *Eur. Phys. J. C*, 84(6):651, 2024.

[16] A. Li, Z. Fu, L. A. Winslow, C. Grant, H. Song, H. Ozaki, I. Shimizu, and A. Takeuchi. KamNet: An integrated spatiotemporal deep neural network for rare event searches in KamLAND-Zen*. *Phys. Rev. C*, 107(1):014323, 2023.

[17] N. Levashko, A. Alexander, V. Biancacci, A. Chernogorov, and A. Li. Determination of Dead Layer Parameters of Semiconductor Germanium Detectors Using Machine Learning for the Legend Experiment. *Physics of Atomic Nuclei*, 86(6):1456–1460, December 2023.

[18] G. Fantini et al. Machine Learning Techniques for Pile-Up Rejection in Cryogenic Calorimeters. *J. Low Temp. Phys.*, 209(5-6):1024–1031, 2022.

[19] Aobo Li, Julieta Gruszko, Brady Bos, Thomas Caldwell, Esteban León, and John Wilkerson. Ad-hoc Pulse Shape Simulation using Cyclic Positional U-Net. In *36th Conference on Neural Information Processing Systems: Workshop on Machine Learning and the Physical Sciences*, 12 2022.

[20] Brendan J. Frey and Delbert Dueck. Clustering by Passing Messages Between Data Points. *Science*, 315(5814):972–976, February 2007. Publisher: American Association for the Advancement of Science.

[21] Corinna Cortes and Vladimir Vapnik. Support-Vector Networks. *Machine Learning*, 20(3):273–

297, September 1995.
- [22] Esteban León. AP-SVM Data Cleaning, September 2024.
- [23] Esteban León. Training and testing data for ap-svm, September 2024.
- [24] N. Abgrall, M. Amman, I.J. Arnquist, et al. The MAJORANA DEMONSTRATOR readout electronics system. *Journal of Instrumentation*, 17(05):T05003, may 2022.
- [25] Michael Willers. Signal Readout Electronics for LEGEND-200. *Journal of Physics: Conference Series*, 1468(1):012113, feb 2020.
- [26] A. Gadola, C. Bauer, F. Eisenkolb, et al. FlashCam: a novel Cherenkov telescope camera with continuous signal digitization. *Journal of Instrumentation*, 10(01):C01014, jan 2015.
- [27] J.F. Wilkerson and Mark Howe. Object-oriented Real-time Control and Acquisition, 2024.
- [28] Matteo Agostini, Jason Detwiler, Luigi Pertoldi, et al. pygama, August 2024.
- [29] Luigi Pertoldi, Jason Detwiler, Sam Borden, Samuel L. Watkins, Christian Nave, James Browning, and Tim Mathew. legend-daq2lh5, May 2024.
- [30] Jason Detwiler, Luigi Pertoldi, Ian Guinn, Grace Song, Sam Borden, Moritz Neuberger, Patrick Krause, and Manuel Huber. legend-pydataobj, August 2024.
- [31] Ian Guinn, Luigi Pertoldi, Jason Detwiler, Sam Borden, Ben Shanks, Clint Wiseman, Tim Mathew, Valerio D'Andrea, Patrick Krause, George Marshall, Matteo Agostini, Gulden Othman, and Esteban León. dspeed, February 2024.
- [32] G.K. Giovanetti, N. Abgrall, E. Aguayo, F.T. Avignone, and A.S. Barabash. A Dark Matter Search with MALBEK. *Physics Procedia*, 61:77–84, 2015. 13th International Conference on Topics in Astroparticle and Underground Physics, TAUP 2013.
- [33] Clint Wiseman and for the MAJORANA Collaboration. A low energy rare event search with the MAJORANA DEMONSTRATOR. *Journal of Physics: Conference Series*, 1468(1):012040, feb 2020.
- [34] Padraic Seamus Finnerty. *A Direct Dark Matter Search with the MAJORANA Low-Background Broad Energy Germanium Detector*. PhD Thesis, University of North Carolina at Chapel Hill, Chapel Hill, NC, May 2013.
- [35] Martin Ester, Hans-Peter Kriegel, Jörg Sander, and Xiaowei Xu. A density-based algorithm for discovering clusters in large spatial databases with noise. In *Proceedings of the Second International Conference on Knowledge Discovery and Data Mining*, KDD'96, page 226–231. AAAI Press, 1996.
- [36] Tian Zhang, Raghu Ramakrishnan, and Miron Livny. Birch: an efficient data clustering method for very large databases. *SIGMOD Rec.*, 25(2):103–114, June 1996.
- [37] T. Cover and P. Hart. Nearest neighbor pattern classification. *IEEE Transactions on Information Theory*, 13(1):21–27, 1967.
- [38] Carl E. Rasmussen and Christopher K. I. Williams. *Gaussian Processes for Machine Learning*. MIT Press, 2006.
- [39] Leo Breiman. Random Forests. *Machine Learning*, 45(1):5–32, October 2001.
- [40] F. Rosenblatt. The perceptron: A probabilistic model for information storage and organization in the brain. *Psychological Review*, 65(6):386–408, 1958. Place: US Publisher: American Psychological Association.
- [41] Thomas Hofmann, Bernhard Schölkopf, and Alexander J. Smola. Kernel Methods in Machine Learning. *The Annals of Statistics*, 36(3), June 2008. arXiv:math/0701907.
- [42] Koby Crammer and Yoram Singer. On the Algorithmic Implementation of Multiclass Kernel-based Vector Machines. *Journal of Machine Learning Research*, 2(Dec):265–292, 2001.
- [43] Laurens van der Maaten and Geoffrey Hinton. Visualizing Data using t-SNE. *Journal of Machine Learning Research*, 9(86):2579–2605, 2008.
- [44] LONGLEAF CLUSTER, 2024.
- [45] F. Pedregosa, G. Varoquaux, A. Gramfort, et al. Scikit-learn: Machine Learning in Python. *Journal of Machine Learning Research*, 12:2825–2830, 2011.
- [46] Olivier Grisel, Andreas Mueller, Lars, et al. scikit-learn/scikit-learn: Scikit-learn 1.3.2, October

2023.
[47] Gina Grünauer. Muon Veto of the LEGEND Experiment. In *Proceedings of XVIII International Conference on Topics in Astroparticle and Underground Physics — PoS(TAUP2023)*, volume 441, page 261, 2024.
[48] M. Agostini, G. Araujo, A. M. Bakalyarov, et al. Pulse shape analysis in GERDA Phase II. *The European Physical Journal C*, 82(4):284, April 2022.
[49] S. I. Alvis, I. J. Arnquist, F. T. Avignone, et al. Multisite event discrimination for the MAJORANA DEMONSTRATOR. *Phys. Rev. C*, 99:065501, Jun 2019.
[50] Clopper C. J. and E. S. Pearson. The use of confidence or fiducial limits illustrated in the case of the binomial. *Biometrika*, 26(4):404–413, 12 1934.
[51] I. J. Arnquist, F. T. Avignone, A. S. Barabash, et al. $\alpha$-event characterization and rejection in point-contact HPGe detectors. *The European Physical Journal C*, 82(3):226, March 2022.
[52] Frank Edzards, Lukas Hauertmann, Iris Abt, et al. Surface characterization of p-type point contact germanium detectors. *Particles*, 4(4):489–511, 2021.
[53] Rosanna Deckert, Igor Abritta, Gabriela Araujo, et al. The LEGEND-200 Liquid Argon Instrumentation: From a simple veto to a full-fledged detector. In *Proceedings of XVIII International Conference on Topics in Astroparticle and Underground Physics — PoS(TAUP2023)*, volume 441, page 256, 2024.
[54] Oliver Schulz, Felix Hagemann, Florian Henkes, Lisa Schlueter, Esteban León, Rosanna Deckert, Arthur Butorev, Ann-Kathrin Schuetz, Andreas Gieb, Danielle Schaper, and David Hervas Aguilar. Juleana.jl, June 2024.